# Various Types of Aesthetic Curves

R.U. Gobithaasan
Department of Mathematics,
Faculty of Science & Technology,
University Malaysia Terengganu
Tel: 609 6683534 Email: gr@umt.edu.mt

**Abstract**

The research on developing planar curves to produce visually pleasing products (ranges from electric appliances to car body design) and indentifying/modifying planar curves for special purposes namely for railway design, highway design and robot trajectories have been progressing since 1970s. The pattern of research in this field of study has branched to five major groups namely curve synthesis, fairing process, improvement in control of natural spiral, construction of new type of planar curves and, natural spiral fitting & approximation techniques. The purpose of is this paper is to briefly review recent progresses in Computer Aided Geometric Design (CAGD) focusing on the topics states above.

## 1 Introduction

Geometric modeling deals with the study of free-form curve and surface design and it is one of the most basic tools in product design environment. The superset of geometric modeling is computational geometry which encompasses computer-based representation, analysis and synthesis of shape information. The latter is now well known as Computer Aided Geometric Design or abbreviated as CAGD. The mathematical entities of product development involve CAGD functionalities which lead to Computer Aided Design (CAD) systems development. Three general phases of product development are creative, conceptual and engineering phase. CAx (Figure 1) is an acronym which represents various IT systems support of all the phases involve in the lifecycle of a product [1]. The definition of CAGD deals with the construction and representation of free-form curves, surfaces and volumes [2]. The abbreviation was first introduced by Barnhill and Riesenfeld for a conference in 1974 held at the University of Utah. The event successfully drawn many researchers from all over the world who were enthusiastic pertaining to the topics involved in computational geometry used for design and manufacture. The event is regarded as the founding event of the new discipline; CAGD. Since then, many conferences and workshops have been held and, text books and journals have been published around US and Europe which contributed to the intensification of CAGD.

Even though differential geometry explains in detail concerning of curves and surfaces, but its potential were not known to Computer Aided Design (CAD) and Computer Aided Manufacturing (CAM) environment. CAD models are mathematically precise geometrical descriptions of a physical objects and the description include numerical data as well as algorithms to prescribe the geometry of the objects [3]. The exploration and detailed investigation of mathematics for CAD/CAM and computer graphics was the precursor to start a brand new field of study called CAGD [2].

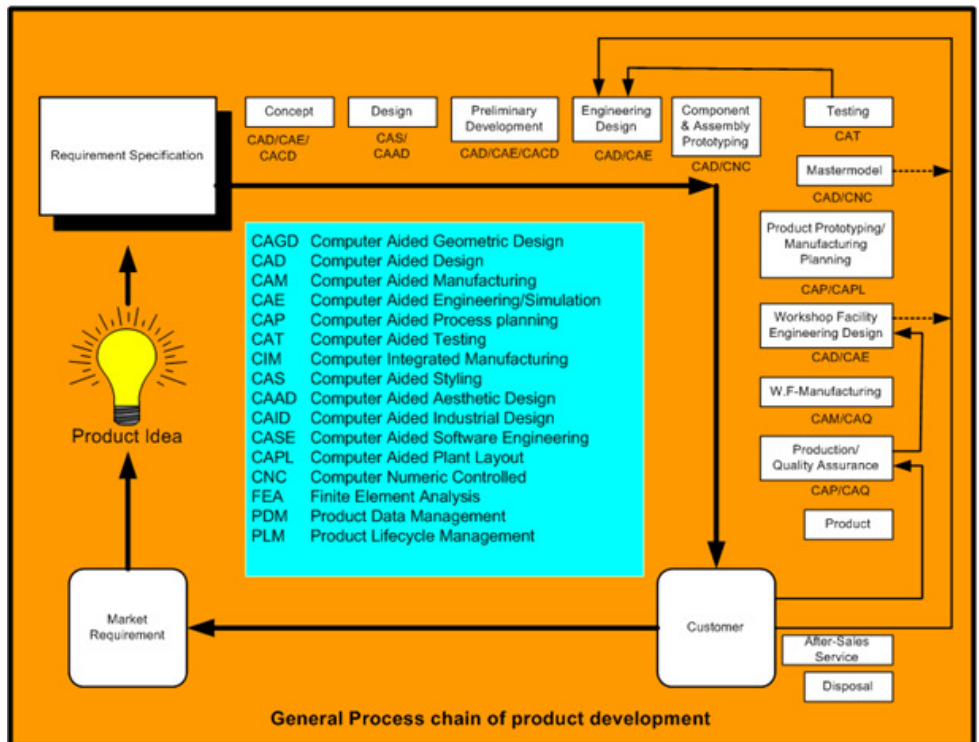

Figure 1: General process chain of product development [1].

The process of extraction of information from geometrical model is known as geometry processing or shape interrogation [4]. Shape interrogation techniques are vital for product manufacturing to verify whether the designed product meets its functionalities and aesthetic shapes. The most widely used shape interrogation technique in 2D environment is curvature profiling whereby the designer reshape the curve by while inspecting the curvature profile or porcupine plot. There are many techniques available for 3D environment namely, contour lines, lines of curvature, asymptotic lines, isophotes and reflection lines [5]. In automobile industry, the skin of the car is investigated using reflection line method. Since the tweaking of shape parameters to reach to a pleasing curve or surface is a painstaking process, researchers use artificial intelligence techniques, for example genetic algorithm to carry out the task [6].

The definition of free form shapes for consumer appliances in terms of the stylists, designers and model makers has been thoroughly investigated in [7]. The term styling and design can be differentiated as follows:

- Styling: concentrates on modeling the outer appearance of a product. It uses emotional and aesthetic values and does not serve any technical or functional purpose.

- Design: is the creative integration of technology, function and form (mathematical formulation and its notation falls under this category).

For more details, see [8, 9,10].

The customer decides to buy a product after having carefully considered four factors namely efficiency, quality, price and its appearance. Pugh [11] has stated that customers judge the aesthetic appeal of a product before the physical performance. This clearly indicates the importance of aesthetic shapes for the success of marketing an industrial product. According to [12], the objects that are manufactured can

be classified into three categories:

- Parts designed based on its technical functionalities and their shapes are determined by rigorous tests and successive approximation. Examples include propellers, aircraft wings, turbine foils, ship hulls etc.

- Parts designed based on its functionalities, often hidden and the only requirement is these parts do not collide with other parts. These parts are often designed in such a way to ease manufacturing process (stamping, forging, casting or machining). Examples are inner panels of electrical appliances, car bodies and etc.

- Parts which are apparent and have to fulfill aesthetic requirements. Examples include the skin of car bodies, household products, sports and leisure equipments, glassware, the outer part of electrical appliances etc.

They further stated that the solution of the third category must be primarily interactive. The reason behind it is stylists with strong intuition about shape but limited knowledge of mathematical models must be able to control complex mathematical representation of curves and surfaces. An example of solution is Unisurf [13] and it has been vastly used by the French car industry. In other words, the stylist and designer must be able to spend more time on the ergonomic shape of the model instead of going through repeated modification of the prototype in order to arrive at an aesthetically appealing shape. Hence, an ultimate system for industrial product design must be geometrically interactive for stylists and designers and the shape interrogation tools must work in hand in order to propose other probable aesthetic shapes.

## 2 The Importance of Curvature Profile & Beyond

The curvature profile for a parametric curve can be obtained by plotting parameter t and against signed curvature value. It indicates how the curvature varies as a point progresses along the curve. The absolute value of the reciprocal of curvature equals to radius of curvature. The area under a curvature profile for a planar curve equals to its turning angle. Thus, similar value of area for a curvature profile can be represented by different shapes of curves. The curvature is curve intrinsic and it is geometric invariant [14]. The shape of the curve is not affected by translation, reflection about an arbitrary axis, rotation and uniform scaling. Even though uniform scaling changes the curvature, the number and relative positions of curvature extrema, cusps, inflection points and self-intersections are preserved [15].

In the field of rapid manufacturing, curvature value determines the preprocess error. It occurs during the conversion of CAD to standard tessellation language (STL) format (as machine input) whereby the outer surface of the part is estimated to triangles. As the value of curvature goes higher, this type of error increases. Even though meshing with smaller triangles may diminish the error, it increases file processing time, and forms more complicated laser trajectory [16].

Planar curves which are visually pleasing has been denoted with many terms, e.g., fair curves, beautiful curves, aesthetic curves, monotonic curvature curves and etc. In this paper, the term aesthetic curve has been used to denote a visually pleasing curve. In differential geometry, a spiral segment means a curve with monotone curvature of constant sign (pg.48,[17]). Farin defined a fair curve as a curve which generates continuous curvature profile and consists of only a few monotonic pieces (pg.364, [14]). Although he admits that the definition of fairness is subjective, this term has been widely accepted to denote visually pleasing curves.

The construction of visually pleasing curves especially planar curves with monotonic curvature profile has been an on-going process. Upon carrying out extensive literature review in this field of research, the classification, based on the type of research, is proposed as follows:

**Curve synthesis**: The idea relies on creating a planar curve from given curvature profile which is somewhat in reverse order than usual.

**Fairing process**: The process of tweaking control points with the help of shape interrogation technique namely curvature profile to obtain the desired curve. The fundamental principle is to reduce the occurrence of wriggles in curvature profile by tweaking control points, polygon or points which control the shape of the curve.

**Improvisation in control of natural spiral**: Modification of planar spirals in order to acquire better control of its shape to suit the curve design environment. To note, past researches are clothoid oriented whereby research on other spirals, for example, Logarithmic spirals and circle involutes are either scarce or sidelined.

**Construction of New Type of Planar Curve**: This type of research involves the manipulation of variables meant for controlling the shape of the curve in order to reach a monotonic curvature profile. Hence, the end product of this research will be a curve with less degree of freedom (DOF) as compared to its original DOFs with monotonic curvature profile.

**Natural spiral fitting and approximation techniques**: The process of approximating natural spirals using flexible curves, namely rational Bezier cubic and B-spline curves falls in this group. The main idea of this field of study is to represent natural spirals (which are usually in transcendental form) in polynomial form so that it can be incorporated into commercial CAD systems.

The following sections discuss in detail the fundamental research papers involved in this field of study based on the stated classification. Selected research papers are thoroughly investigated and elaborated.

## 3  Curve Synthesis

In 1970s, the CAD research group at the Engineering department of Cambridge University investigated the techniques involved for integrating curvature profiles for planar curves. The idea behind this method is to construct a curve based on given curvature profile. The following principal cases have been formulated during1970s:

- $\kappa(t)=0$; the curve generated is a straight line.
- $\kappa(t)=$constant$=1/r$; a circle of radius of r is generated.
- An abrupt step change in curvature from zero to constant value; a straight line connected to a circular arc.
- An abrupt step change in curvature from one constant $\kappa(t)\neq 0$ to another constant
- $\kappa(t)\neq 0$; the generation of two circular arcs connected with different radii.
- $\kappa(t)= a\,t$; A segment of clothoid is generated with a as a constant.

The extension of [18] is [19], in which an algorithm to generate a smooth curve with given data points, curvature and tangent direction at those points. Again, clothoid has been used for curve synthesis purpose. A complete package has been proposed for generation of linear curvature spline package for general use of curve synthesis in [19].

A curvature function which is equivalent to a constant value is the simplest form of curvature function and it produces a segment of a circular arc. A linear curvature function or denoted as LINCE represented by $\kappa(s)=a+bs$ where a and b are constants; it corresponds to a segment of a clothoid. Figure

2 illustrates an example of LINCE and its corresponding curve. Similarly, a quadratic curvature function defined by κ(s)=as²+bs+c (where a, b and c are constants). Since oscillations may occur in quadratic curvature function, hence it is not preferable. However, a quadratic curvature function cannot be reduced to Logarithmic spiral [20]. A bilinear curvature function or denoted as BLINCE is a curvature function in the form of rational linear function and it generates Generalized Cornu Spiral (GCS).

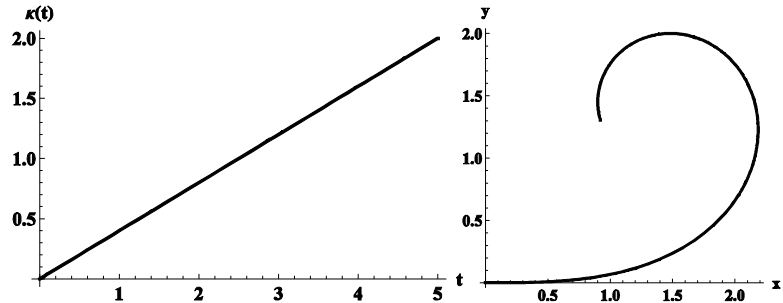

Figure 2: LINCE (left) and its corresponding curve.

## 4  Fairing Methods

In general, there is a process called fairing technique (smoothing denotes similar process but it is used widely for digitized point data) involved in designing visually pleasing curves and surfaces. Example of products are the roof of a car in which undesirable wriggles must be avoided and the wing or tail of aircrafts must be oscillation free in order to meet the aerodynamic properties. The figure below illustrates stages that are involved in a fairing process [21].

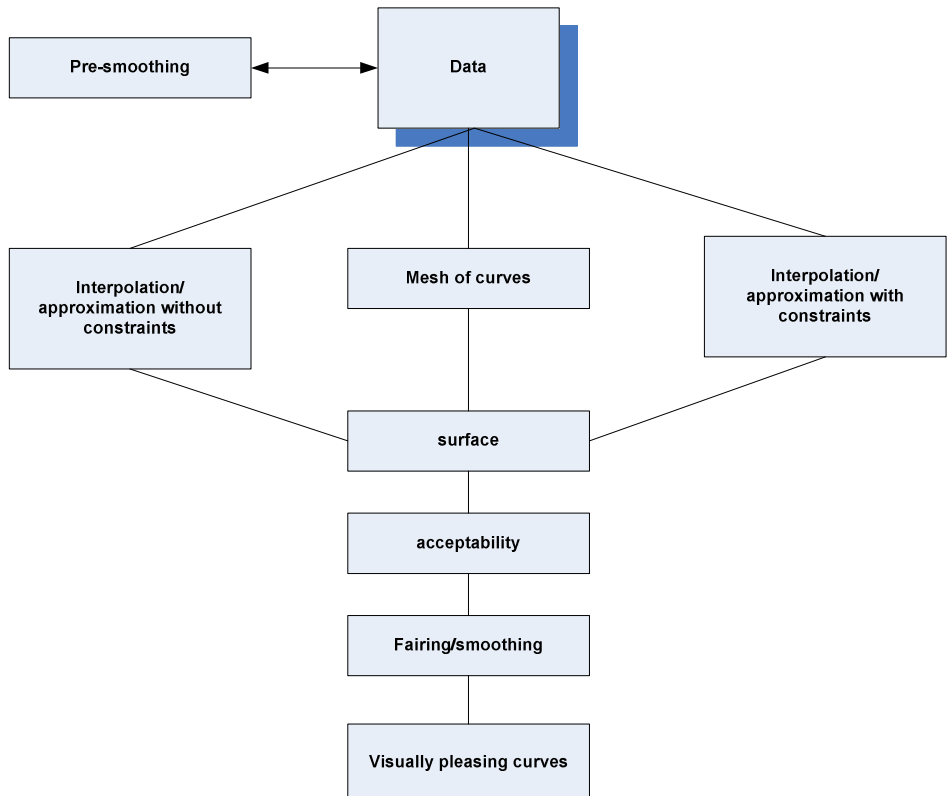

Figure 3: Curve and surface modeling via fairing process [21].

Pre-smoothing technique for digitized point data has been extensively discussed in [22] whereby the proposed technique is based on using difference quotients to discretize various geometric invariants namely curvature values which is proportional to the second derivative. Typically there is a step which may involve some form of bending energy in order to obtain a visually pleasing curve, for example minimal bending ($\int_a^b |f'''| dt$), minimal bending energy ($E = \int_a^b \kappa^2 \, ds$) and etc (refer to [21] and references therein for more techniques).

The simplest way to obtain a visually pleasing curve is by interactively modifying its control points based on curvature profile. The shape interrogation techniques are utilized to identify bad curve segments which is later improved via fairing process. According to Farin (pg.364, [14]), a planar curve with almost piecewise linear is said to be fair. Hence, the continuity of curvature profile is an essential indicator for the aesthetic value of an arbitrary shape. Many algorithmic methods are available for the stated matter, in [23, 24, 25].

A criteria on reducing the difference in curvature between biarc curves has been discussed in [26] which results with a global fairing technique. A method to fair composite polynomial curves with approximation conditions, end conditions and integral conditions (the area under the curve) has been investigated in [27]. More research on identifying bad data points and replacing them with new ones in order to derive a visually pleasing shape can be found in [28,29]. A method which approximates to targeted curvature profile for B-Spline by minimizing the energy versus coherence to the original curve design has been proposed in [30].

Another interesting work on curve fairing is by Yamada et.al [31]; whereby the fairing process model was developed based on discrete spring model. The imaginary spring is attached along the normal line of each node. The length of the spring represents the curvature value and the energy represents the difference of the lengths of neighboring springs , hence the iterations are carried out until the total energy is minimized.

A notable work on faring a piecewise curve can be found in [32], whereby Miura et.al proposed a discrete log-aesthetic filter to carry out the fairing process. This filter greatly reduces the number of iterations involved to produce a visually pleasing curve.

## 5 Improvisation in Control of Natural Spiral

There are many studies which indicate the use of natural spirals for the design environment. The main reason for using natural spirals namely clothoid, logarithmic spiral and circle involute is that its
curvature formula is simple unlike the curvature formula of conventional curves, e.g. Bezier and B-spline curves. However, natural spirals consist of transcendental functions; the formulation which involves either integration function or sine and cosine function. To add, the ability to control these kind of curves has always been an issue for the CAD environment. Since the development and advancement of computation power are exceptionally promising, the generation of these kind of spirals are almost effortless. Hence, the focus has been on developing algorithms to ease the controllability of the shape of spirals to suit the intent of the design.

Clothoid has been used for highway design for a long time since it has the property that its curvature varies linearly with arc length. Bass has introduced five clothoid templates to design highways [33]. His paper has been a pioneer for the investigations of controlling clothoid that can be found in [34-39]. The investigation on finding the offset of clothoid can be found in [40]. A patent on calculating shape parameter to determine the size of a clothoid that needs to be inserted when designing a S-shaped and interchange, and a connection road can be found in [41]. The usage of clothoid curve (referred as 3D curve) for designing motion trajectories of mechanical elements can be found in [42].

There are less work concentrated in other types of spirals regardless of the monotonicity of their curvature profile such as the Logarithmic spiral and circle involute. In 1998, Mitsuru et.al [43] proposed a method to construct piecewise Logarithmic spiral with $C^2$ continuity. After a year, Mitsuru et.al [44] proposed on designing piecewise circle involute in $G^1$, $G^2$ and $G^3$ terms.

## 6  Natural Spiral Fitting and Approximation Techniques

Another way of developing a visually pleasing curve is by fitting spirals to a given data sets. This field of study is well known as curve fitting problem. However, earlier works on designing visually pleasing curves involve biarcs, even though these type of curves can only result with tangential continuity ($G^1$). This is because the resultant piecewise curve reduces the wriggles that occur in the curvature profile unlike the conventional curves such as B-spline curves. In 1990s, biarcs are available in NC machines thus many algorithms have been developed for curve fitting with biarcs, e.g., Schonherr [45] developed a biarc curve fitting technique which minimizes the difference in the curvatures of two adjoining circular arcs.

Since the development of piecewise curve with curvature continuity ($G^2$) is far more essential, extensive amount of work has been proposed. In 1974, Mehlum published a work on approximating clothoid using a sequence of stepped circular arcs of linearly increasing curvature [46]. Stoer [47] improvised Mehlum's work by minimizing $\int_a^b \kappa^2\, ds$ among all interpolating clothoidal splines and he showed that his technique of clothoid approximation is of high quality whereby the deviation from the given points is not greater than a prescribed tolerance.

Another example of article related with this field of study is [48], whereby Davis used Total Least Square (TLS) to directly fit clothoid to given measured points. In 2000, Wang et.al [49] proposed an algorithm to approximate Clothoid with Bezier and B-spline in the interval of [0,Π/2]. Sanchez and Chacon proposed polynomial approximation to clothoids via s-power series and they claimed that their technique approximates better than Wang et.al's technique.

In 2004, Meek & Walton [38] proposed a technique to approximate clothoid using a set of arc splines. The selected piecewise clothoid is then converted into discrete clothoid whereby each segment is represented with an arc spline. Recently, Real-time clothoid approximation using Rational Bezier curves was proposed in [50].

Other spirals that have been approximated in polynomial form for CAD system incorporation are Logarithmic spiral and circle involute. In 1997, Baumgarten & Farin proposed an algorithm to approximate Logarithmic spiral with rational cubic spline curve [51]. In 2007, Higuchi et. al used Chebyshev approximation formula [52] to approximate circle involute in terms of polynomial which enable them to represent circle involute as a Bezier curve.

## 7  Construction of New Types of Planar Curve

There are many types of curves which have been constructed with the aim of designing visually pleasing shapes. An ideal curve must be flexible enough for the stylists to model a product or shape in accord to his/her desire similar to a function of pencil or a pen. Examples of traditional curves suitable for free form design are Bezier and NURBS. The designer and CAD specialists must then be able to regenerate the model proposed by the stylists in a CADCAM system for manufacturing without destroying the principal shape characteristics of the model. This step would involve the shape interrogation techniques namely curvature profile. Hence, the curvature profile of curves used for modeling must be piecewise linear as proposed by Farin [14]. If the stylists uses traditional curves for modeling, then an extra step is needed for fairing the model, and this is painstaking process. Thus, mathematician and design engineers investigated on developing curvature controlled curves. This led to the development of many new types of curves which are elaborated in the following sections.

## 7.1 Bezier Spiral

Planar Bezier cubic has eight degrees of freedom by nature; two degrees of freedom at each control points. Meek & Walton have contributed significantly in restricting the degree of freedom of this curves to suit the design environment and the restricted curve is denoted as cubic Bezier spiral [53]. It is an example of a curve which has monotonic curvature profile. The proposed curve has five degrees of freedom available for design intent, specifically two degrees of freedom for the first control point $B_0$, g=h, k and $\phi$. In order to obtain planar cubic Bezier spiral, the configurations are set in such a way that the first three control points must be collinear and the second control point is the midpoint of the first and third control point as shown in Figure 4.

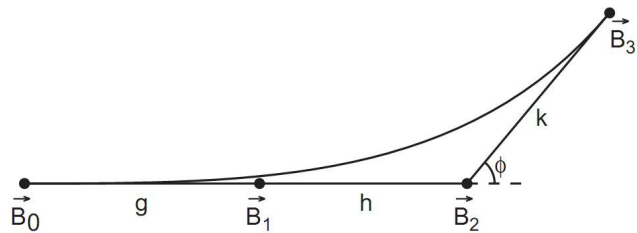

Figure 4: Configuration of control points to obtain a monotonic curvature profile as proposed by Walton & Meek [53].

Yeoh & Ali [54] have extended this work in which they added an extra degree of freedom by letting the second control point to be anywhere on the line connecting the first and third control point. In the same year, Walton et.al proposed the generalization of cubic Bezier spiral curve with the configuration shown in Figure 5 [55].

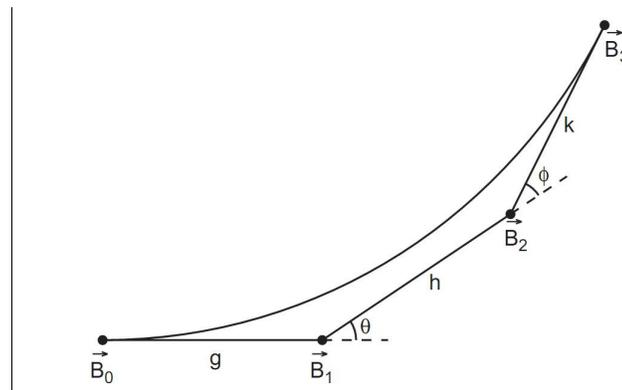

Figure 5: Configuration of control points to obtain a monotonic curvature profile as proposed by Walton et.al [55].

Similar restriction procedures have been adopted by Azhar et.al [56] to develop quartic Bezier spiral whereby it has seven degree of freedoms and it is much more flexible as compared to cubic Bezier spiral. Hence, a designer has more choices to shape the desired curve. Since Bezier spiral loses its flexibility, the family of these kind of curves are small in numbers. Furthermore, the variables used to determine the shape of the curve has many conditions to meet in order to generate a visually pleasing curve. The selection of right control points to maintain Bezier curves with monotonic curvature profile, specifically parametric cubic spiral which interpolates given positional and tangential end conditions has been studied by Dietz & Piper [57]. They further extended their work to develop rational cubic spirals [58].

Regardless of the restriction of variables of Bezier curve to obtain monotonic curvature profile, the linearity of the curvature profile cannot be as smooth as clothoid's curvature profile. Harada et.al has shown that it is possible to approximate cubic Bezier spiral to clothoid via Logarithmic Distribution of Logarithmic Distribution Diagram of Curvature or LDDC [59].

The conditions for general rational quadratic B-spline curves with any number of control points to be spiral has been derived in [60] and in [61] different kind of fairing process is introduced whereby the detailed function is based on a Multi-Level Analysis instead of calculating the energy integrals or curvature derivatives of cubic B-spline curves.

## 7.2 Pythagorean Hodographs

Pythagorean Hodographs or PH curves was first introduced by Farouki and Sakkalis [62]. There are two features that makes this curve interesting that are its arc length which can be expressed as a polynomial function of the parameter and the formula for its offsets is a rational algebraic expression. The lowest degree of PH curves that contain inflection points are PH quintics and this type of curve has eight degrees of freedom. Walton & Meek [63] proposed PH quintic which has curvature that varies monotonically with its arc length, denoted as Pythagorean Hodographs quintic Bezier spiral or in short H-spiral. This curve has five degree of freedoms. However, the usage of the algorithm involves the incorporation of straight lines and circular arcs when curvature extrema are dealt with. Thus, PH spiral is proposed as a transition curve between straight lines and circular arcs similar to the case of cubic Bezier spiral. Farouki [64] improved the result by representing PH quintic a in complex form whereby the new result gives a compact derivation and reveals the full shape flexibility of PH quintic spirals. Later, Walton & Meek [65] improved further the work on designing curves using PH spiral and concluded that PH spirals are much more flexible than cubic Bezier spiral and this curve was proposed to substitute clothoid wherever the curvature controlled curves are needed. This new generalization maximizes the degrees of freedoms available to render composite PH spirals with $G^2$ continuity.

In 1996, Farouki et.al [66] has developed Real-time CNC interpolators for PH curves and further investigated on feedrate and material removal rates which is reported in [67]. Other recent work on PH curve include [68-70].

## 7.3 Related Curvature Controlled Curves

Many other types of planar curves developed with the aim of generating visually pleasing curves include Tschirnhausen cubic spirals[71], unit quaternion integral curves [72], Monotone Curvature Variation (MCV) curves [73], Minimum Variation Curves (MVC) via Sobolev gradient method [74] and optimized geometric Hermite curves (OGH) [75]. Miura [76] proposed a new type of curve called log-aesthetic curves. This curve was the result of the extension of Harada et.al's work on LDDC [77]. The advancement of this curve seems promising.

Class A Bezier curve was first coined by Farin [78] after the term "class A surface" used in Mercedes-Benz CAD/CAM system (SYRKO) which means highest grade of shape quality. This 3D Bezier curve is the resultant of the generalization of certain 2D Bezier curves (referred as typical curve) introduced by Mineur [79] and it has two special properties; the monotonicity of curvature and torsion is preserved. However, under certain conditions, the monotonicity of the stated properties are violated; Cao & Wang have shown the properties and conditions to preserve the monotonicity of curvature and torsion in [80].

# Conclusion

As the computing power increases and its price decreases, we believe that the formulation of visually pleasing curves which involves double integration or sine/cosine function is much more feasible in practice. At the same time, artificial intelligence algorithms can also be incorporated for a better solution.

**Acknowledgments**

This work was supported in part by UMT (SB-Ph.D:53027) and MOHE (FRGS 59187). The author acknowledges Dept of Mathematics, University Malaysia Terengganu for their financial support to present and publish this paper.